\documentclass[aip,apl,amsmath,amssymb,reprint]{revtex4-1}

\usepackage{graphicx}
\usepackage{dcolumn}
\usepackage{bm}

\begin{document}

\title{Directional terahertz emission from diffusion-engineered InAs structures}

\author{Jong-Hyuk \surname{Yim}}
\affiliation{School of Information and Communications, Gwangju Institute of Science and Technology,
Gwangju 500-712, South Korea}

\author{Hoonil \surname{Jeong}} 
\affiliation{School of Information and Communications, Gwangju Institute of
Science and Technology, Gwangju 500-712, South Korea}

\author{Jin-Dong \surname{Song}}
\affiliation{Nano-Photonics Research Center, Korea Institute of Science and Technology, Seoul
136-791, South Korea}

\author{Young-Dahl \surname{Jho}}
\email{jho@gist.ac.kr} \affiliation{School of Information and Communications, Gwangju Institute of
Science and Technology, Gwangju 500-712, South Korea}

\date{\today}

\begin{abstract}
We have designed and fabricated a new type of terahertz (THz) emitter that radiates THz waves along the surface-normal direction because of the lateral distributions of the transient electric dipoles. The excitation and measurements were performed using a conventional THz time-domain spectroscopy scheme with femtosecond optical pulses. The corrugated mirror patterns on the InAs layers made the radiation directional along the surface-normal direction, and the emission efficiency was controlled by adjustment of the pattern width.
\end{abstract}


\maketitle

As an important practical source of pulsed electromagnetic radiation at terahertz (THz)
frequencies, the photo-excitation of semiconductor surfaces by femtosecond (fs) laser pulses has been
widely used over the past two decades. A vast array of compound semiconductors have been examined
with regard to their material parameters, including InAs, InSb, GaAs and InP.\cite{Sakai} The THz radiation
emitted from these materials has been attributed to either photo-carrier acceleration caused by band-bending near the
semiconductor surface or to the large diffusion velocity difference between electrons and holes (the photo-Dember effect). These drift and diffusive transport mechanisms of the photo-excited carriers are
identified separately from those of the different bulk compound semiconductors: In the case of GaAs, the THz
radiation is known to be produced mainly by the surface electric field,\cite{Rice} whereas the
radiation from InAs is mainly governed by the photo-Dember effect.\cite{Gu}
Comparison studies were also performed on the THz radiation from GaAs and InAs,\cite{Liu1} while a
detailed explanation of the THz radiation processes from InAs was provided in terms of drift and
diffusion.\cite{Liu2, Jeong}

Further technological advances have been made toward efficient radiation sources with higher
output powers and/or tunability\cite{Tonouchi} and with better spatial resolution that is sometimes beyond
the diffraction limit.\cite{Wachter} Recently, THz waves with increased amplitude and bandwidth have been reported 
in periodically metal-patterned In$_{0.53}$Ga$_{0.47}$As (and GaAs) by breaking lateral symmetry in
their diffusion currents,\cite{Klatt1, Klatt2} while the transfer of THz waves over
long distances was demonstrated by using either a laser-plasma filament\cite{Amico} or an optical
fiber coupled with tilted InAs tips, while retaining superior spatial resolution.\cite{Yi}

From the materials viewpoint, InAs is known to be one of the most intense sources among
the various candidate semiconductors because it has an electron mobility (reaching up to 30,000
cm$^{2}$/V$\cdot$s) that is much higher than the hole mobility ($\sim$240 cm$^{2}$/V$\cdot$s), leading to efficient
photo-Dember current generation. Also, the 800 nm excitation at the central wavelength of
Ti:sapphire-based laser technologies provides a large photo-Dember field because of the short
absorption depth ($\sim$140 nm) and large excess energy ($\sim$1.2 eV), when considering the narrow
band-gap ($E_{g}=0.35$ eV at 300 K).\cite{Que} Despite the moderate efficiency of InAs-based
transmitters, significant obstacles to the material’s use remain, particularly with regard to the control of the propagation direction
of the THz waves. In conventional THz time-domain spectroscopy (THz-TDS) systems with fs laser
illumination, for example, InAs sources are tilted with an angle of 45$^0$ to the incident beam to be
coordinated with the guiding components; the generated THz waves are therefore inevitably spatially
dispersed. We also note that the diffusion-controlled currents in InAs, with a larger photo-Dember
field than in any other semiconductor, could offer THz emitters that are aligned in a line-of-sight configuration for 
prospective THz applications in imaging and communications. In this work, we have implemented
micro-scale groove patterns in InAs-based structures, with the aim of producing enhanced directional emission of THz
waves, as measured using various detection geometries.

The photo-Dember currents feature prominently along the [100] growth direction
of conventional InAs epilayers, as shown in Fig. 1(a). The 1 $\mu$m-thick InAs epilayers used here were processed to have
grating patterns, acting either as 45$^0$ reflectors with a gap width $W$ (as shown in Fig. 1(b))
or as parabolic apertures with $W$ of 2.5 $\mu$m. The grooves in Fig. 1(b) were proposed to produce 
enhanced lateral diffusion by generation of abrupt photo-carrier gradients at the air-(010) interfaces,
together with the enhanced THz radiation along the surface-normal direction, as depicted by a large
green cone along the [100] direction. The grating grid (with $W$=0.5, 1.2 $\mu$m, or 2.5 $\mu$m) was
fabricated by the shot-modulation technique and dry etching on InAs epitaxial layers (containing $p$-type doping of $1.5
\times 10^{19}$ cm$^{-3}$ with $Be$) grown on 500 $\mu$m-thick GaAs substrates; scanning electron microscopy (SEM) images of the structure are
shown in Fig. 1(c). A 2.5 $\mu$m-thick GaSb layer was inserted to compensate for the lattice mismatch
between InAs and GaAs. The gap width $W$ was designed to have a similar scale to the diffusion
length of the electrons along the vertical [100] direction ($\sim$ 1 $\mu$m).\cite{Jeong} We
performed THz-TDS measurements under excitation with a pulsed Ti:sapphire laser at 300 K (pulse duration
$\sim$ 150 fs, centered at 800 nm). The incident angle $\theta$ of the IR beam was either 45$^0$ in
the conventional reflective detection geometry or 90$^0$. When $\theta$=90$^0$, for comparison
purposes, the photo-conductive antenna (PCA) was either placed along the IR laser in the normal
detection geometry or along the surface in the lateral detection geometry. The pump beam supplies an
excitation fluence of about 0.9 $\mu$J/cm$^{2}$ on a 800 $\mu$m beam diameter; under this regime, the
optical rectification is negligible when compared to the photo-Dember effect.\cite{Reid}

\begin{figure}[!t]
\centering
\includegraphics[scale=0.32,trim=25 0 0 0]{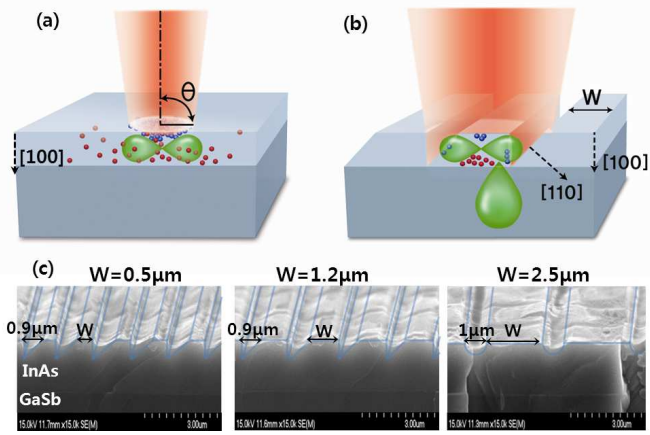}
\caption{Illustrations of the THz generation processes (a) without and (b) with the groove patterns. The incident IR laser beams (shaded red cone) are assumed to be normal to the surface ($\theta$=90$\,^{\circ}$), and the diffusion directions of the photo-generated electrons (blue dots) and holes (red dots) determine the radiation patterns (green dumbbell-like cones) accordingly. (c) SEM images of the
fabricated patterns with $W$ of 0.5 $\mu$m, 1.2 $\mu$m and 2.5 $\mu$m. }
\end{figure}

\begin{figure}[!t]
\centering
\includegraphics[scale=0.3,trim=30 50 0 80]{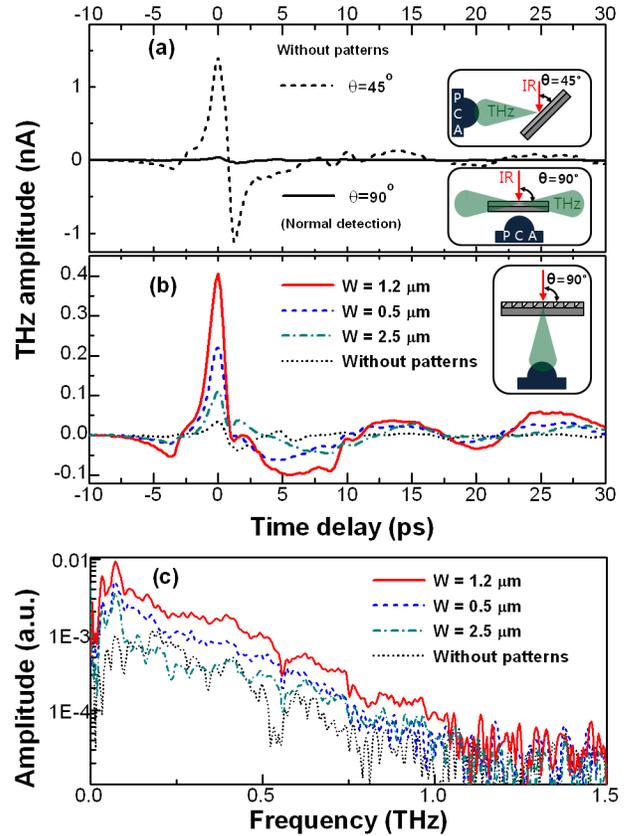}

\caption{Time domain THz measurements of (a) a bare InAs layer at different incident angles ($\theta$=90$\,^{\circ}$ or $\theta$=45$\,^{\circ}$), as shown in the insets, and (b) InAs layers with different groove pattern widths in a transmissive normal detection geometry. (c) Fourier transform spectra of Fig. 2(b).}
\end{figure}

\begin{figure}[!t]
\centering
\includegraphics[scale=0.35,trim=0 50 0 70]{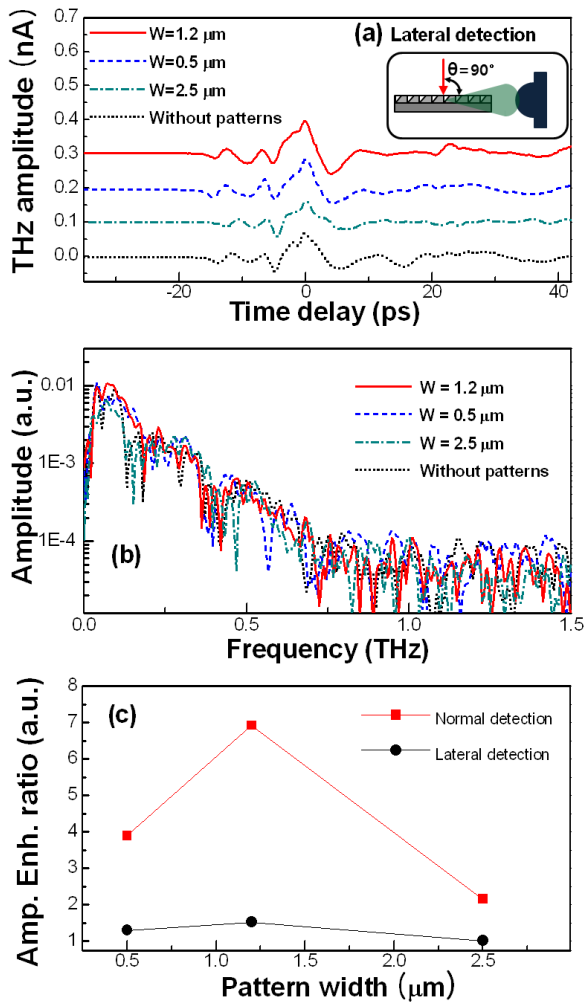}

\caption{THz measurements of (a) InAs layers with and without groove patterns in the lateral detection geometry. (b) Fourier transformed spectra. (c) Emission amplitude enhancement ratios for comparison of the patterned samples to the bare sample for both detection geometries.}
\end{figure}

THz-TDS measurements, similar to those conducted previously in InAs epilayers without the groove
structures, were conducted in as shown in Fig. 2(a), in both the conventional $45^0$ reflection geometry and in
the surface-normal transmissive geometry (or normal detection geometry). The experimental setups
used for the two geometries are depicted in the inset figures. The emission along the surface-normal
direction (solid line) in Fig. 2(a) was relatively suppressed, which can be attributed to the
combined effects of longitudinal diffusion and the reduced radiation out-coupling efficiency caused by the
refractive index mismatch at the interfaces.\cite{Johnston} The slight and slow oscillations shown in Fig.
2(a) are related to the Fabry-Perot effect, for which the period of 12.5 ps is well matched with the round-trip
time in the GaAs substrate.

The lateral photo-carrier density modulation for surface-normal incidence could remove a major
hindrance that affected the transmissive emission suppression, as shown in Fig. 2(a). When the wedge-shaped
groove patterns were configured for lateral diffusion, as shown in Fig. 2 (b), and the IR pump beam
was orthogonally polarized relative to the groove axis, the THz amplitude increased with evident pattern
scale ($W$) dependence. The results from these patterned structures were compared to the radiation from
a bare sample (short dotted line), indicating the enhanced lateral diffusion. The strongest
amplitude was observed in a sample with $W$ of 1.2 $\mu$m, which was similar to the reported value
of the maximum diffusion length of 1.3 $\mu$m \cite{Que} and could possibly be associated with the
optimized lateral diffusion. On the other hand, the out-coupling efficiency in the
groove-patterned samples could be similar to that in the epilayer, as inferred from the persistently
observed Fabry-Perot oscillations. The ratios of the peak amplitudes to the Fabry-Perot peaks were almost
the same among the samples shown in Fig. 2(b). These results are somewhat intuitive, because the wavelength of the THz
waves is much larger than the groove scale, such that we would not expect enhanced scattering
at the interfaces of the patterned samples along the surface-normal direction. In Fig. 2(c),
the Fourier-transformed spectra were taken from Fig. 2(b). In the patterned samples, the lower
frequency side was found to be more greatly enhanced when compared with the bare sample, which is not clearly understood at
this moment. However, there are a few possible examinable points to show the difference: (1) The surface
normal emission in patterned samples could also be affected by the diffusion along the [110] direction with
a higher electron effective mass,\cite{Saidi} thus leading to reduced carrier mobility in the Einstein
diffusion equation of $\mu$ = $e\tau_m$/$m^{*}$, where $m^{*}$ is the effective mass, and $\tau_m$
is the scattering time (cf. Fig. 1(b)). The polar radiation patterns therein from the high density
ensemble of dipoles diffusing along the [110] direction could encompass the emissions along the surface-normal
direction.\cite{Johnston} (2) The delayed, scattered and chirped IR beam \cite{Izumida} in the case
of the patterned samples via the groove patterns could possibly be associated with the low THz band
enhancements, because the IR beam should penetrate into the 1 $\mu$m-scale wedges, in contrast to the
superficial absorption in the bare sample at a scale of 100 nm. The scattering in the IR range could
restrict further emission amplitude enhancements, leading to further complexities in
guiding the THz waves. The spectral dips in Fig. 2(c) are ascribed to absorption by water
vapor.\cite{Roggenbuck}

Lateral emission measurements, under the same IR pumping conditions as those described for Fig.
2(b), were taken on the groove-patterned samples and on a bare epilayer without patterning
and the results are shown in Fig. 3(a) for the time-domain transients and in Fig. 3(b) for the Fourier-transformed spectra.
The geometrical schematic for the lateral detection setup is depicted in the inset of Fig. 3(a), where the
focal spot position of the IR pump beam was displaced by about 2 mm and the Si lens coupled
with the PCA was 1 mm away from the sample edge. The typical time-domain traces in
the lateral detection geometry did not show any distinct features among the samples. The similarity of these
traces implies that the lateral emission is mostly influenced by the propagation
effect via the GaAs layers, which is also manifested in the narrower spectrum shown in Fig. 3(b) that was observed in
waveguide structures with attenuation and dispersion.\cite{Nagel} Also, the comparable emission
amplitude in the lateral detection geometry when compared to the transmissive geometry implies that the
diffusion along the growth direction is still efficient, irrespective of the pattern fabrication. The
amplitude enhancement ratio in Fig. 3(c) was derived from the THz signal intensities, either in the normal detection
geometry or in the lateral detection geometry, and normalized relative to those taken from the bare sample. The
amplitude enhancement ratio from the sample with $W$ of 1.2 $\mu$m was almost twice that of the
sample with $W$ of 0.5 $\mu$m, whereas the values from a sample with periodic apertures ($W$=2.5
$\mu$m) were not significantly different to those from the bare sample. The scale and the structural
dependence of these amplitude enhancements could lead to opportunities for both amplitude and
directional controllability. The optimally adjusted doping concentrations \cite{Adomaviius} 
and improved groove shapes for the effective role of the micro-scale mirrors could further
enhance these emissions along the line-of-sight.

In conclusion, we have developed a new method of implementing groove patterns to act as reflectors
for incident laser pulses in the IR range. The amplitude of the associated THz waves along the surface-normal
direction was alternated, depending on the period and shape of the groove patterns. The
mirror-like patterns that were separated by a gap width of 1.2 $\mu$m were found to be the most efficient among
our samples, whereas periodic aperture patterns with a gap width of 2.5 $\mu$m did not assist
with the directionality when compared to the bare bulk sample. The lateral emission signals, in 
contrast, did not show many differences among the samples, which means that the longitudinal diffusion was
persistent with or without the groove patterns. These results show that, under excitation with IR laser
pulses, the optimization and controllability of the lateral carrier diffusion in InAs could be
useful for alignment-free THz applications in imaging and communications. The groove
emission patterns could possibly be optimized by using low-cost methodologies such as
echelon gratings \cite{Echelon} and pattern imprinting.\cite{Ko}

The authors would like to thank J. Ahn at KAIST for helpful discussions. This work was supported by the
Bio-Imaging Research Center at GIST and by the Basic Science Research Program through the National
Research Foundation of Korea (NRF-2009-0090559). 
The work at KIST was supported by the KIST Institutional Program including Dream Project, MEST 2012K001280 and the Korea-Sweden Research Cooperation Program.

\end{document}